\documentclass[12pt]{article} \input epsf.tex

\newcommand{\be}{\begin{equation}}
\newcommand{\ee}{\end{equation}}
\newcommand{\bea}{\begin{eqnarray}}
\newcommand{\eea}{\end{eqnarray}} 
\newcommand{\ba}{\begin{array}}
\newcommand{\ea}{\end{array}}

\begin{document}

\begin{titlepage}
\def\thepage {}        

\title{Renormalization of the neutrino mass matrix}

\author{
T. K. Kuo$^1$, James Pantaleone$^2$, Guo-Hong Wu$^3$ \\ \\ 
{\small {\it
$^1$Department of Physics, Purdue University, West Lafayette, IN 47907
\thanks{Email: tkkuo@physics.purdue.edu, jim@neutrino.phys.uaa.alaska.edu,
gwu@darkwing.uoregon.edu}}}\\
{\small {\it $^2$Department of Physics, University of Alaska, Anchorage,
 Alaska 99508, }}\\
{\small {\it $^3$Institute of Theoretical Science, University of Oregon,
Eugene, OR 97403}}\\ 
}

\date{\small (Revised, August 2001)}   \maketitle

\begin{abstract}
The renormalization group equations for the general $2\times 2$, complex, 
neutrino mass matrix are shown to have exact, analytic solutions. 
Simple formulas are given for the physical mixing angle, complex phase
and mass ratio in terms of their initial values and the energy scales.
We also establish a (complex) renormalization invariant relating 
these parameters.
The qualitative features of the physical parameters' renormalization 
are clearly illustrated in vector field plots.  
In both the SM and MSSM, maximal mixing is a saddle point.

\end{abstract}

\vfill \end{titlepage}

\section{Introduction}

 Recent experiments \cite{atmos,solar} have revealed some salient features 
of the
neutrino mass matrix. It is now rather well-established that the neutrino
masses are tiny, and that at least some of the mixing angles are 
large, or even maximal. Considerable efforts have been devoted to a 
theoretical understanding of these features. 
Any theoretical model faces an important issue. 
Is the model valid at the high or
low energy scale? How are the physical parameters related between the 
two scales? These questions are answered by studying the renormalization
group equation (RGE). In addition, RGE may offer a natural mechanism
which can drive the neutrino mixing angle to its maximal value.

   The RGEs governing the neutrino mass matrix were worked out some time
ago \cite{babu} and have been studied extensively 
(see e.g. \cite{Casas,other}).
 Most studies considered the case of real matrices, although some
dealt with complex ones also. 
In this paper, we wish to point out that there is
an exact, analytic solution for the RGE of the general 
$2\times 2$ mass matrix. 
Simple formulas are given for the physical mixing angle, complex phase
and mass ratio in terms of their initial values at an energy scale.
We further show that there is a (complex)
RGE invariant relating the running of these three variables.
The fixed points of the physical parameters are obtained
and their stability is determined. 
The nature of the running near these fixed points is 
clearly illustrated in vector field plots \cite{strogatz}.

  In both the standard model (SM) and the minimal supersymmetric standard 
model (MSSM), the effective Majorana neutrino mass matrix can arise from a 
dimension-five operator. Its RGE
(in the basis where the charged-lepton mass matrix is diagonal) is
given by ($t={1\over 16\pi^2} \ln\mu/M_X$),
\be \label{eq:RG} 
{d\over dt} m_\nu = - (\kappa m_\nu + m_\nu P + P^T m_\nu) \ , 
\ee where $\kappa$ depends  on a combination of the coupling constants and, 
for two flavors (for definiteness, we consider $\nu_\mu$ and $\nu_\tau$)
\cite{babu}, after absorbing the muon Yukawa coupling term into $\kappa$,
\be P = P^T = \chi (1-\sigma_3) \ , 
\\ 
\ee 
\be
\chi= \left\{\ba{cl} (y_\tau^2-y_\mu^2)/4 \ , & {\rm SM} \\
   - (\tilde{y}_\tau^2-\tilde{y}_\mu^2)/2 \ , & {\rm  MSSM} \ea \right .
\ee
Here $y_\tau=\sqrt{2}m_\tau/v$ and 
$\tilde{y}_\tau=\sqrt{2}m_\tau/(v\cos \beta)$
are the $\tau$ Yukawa couplings in the SM and MSSM respectively,
with $v\simeq 246 \ {\rm GEV}$ and $\tan \beta$ is given by 
the ratio of the two Higgs VEVs in MSSM. 
Also,  $y_\mu$ and $\tilde{y}_\mu$ are similarly defined.

\section{Solution}

A formal solution \cite{other} to Eq.~(\ref{eq:RG}) is 
\be
m_\nu(t) = e^{-\kappa^\prime t} e^{\xi \sigma_3} m_\nu(0) e^{\xi \sigma_3}
\label{eq:mnu}
\ee
where we have ignored the $t$-dependence of the coupling constants\footnote{
To include the $t$-dependence, one needs only to replace $\kappa t$ by 
$\int \kappa dt$ etc, in the appropriate formulae in the following. Numerically,
we have verified that the difference is negligible.}
 so that
$\int \kappa dt \simeq \kappa t$, {\it etc.}, and
\bea  
\kappa^\prime &=& \kappa + 2 \chi  \\
\xi      &  = & \chi t \ .
\eea
 It is convenient to factor out the determinant
\be   m_\nu=\sqrt{m_1m_2} M \ ,
\ee
then $\det M = +1$, and the mixing angle and the complex mass ratio are
contained solely in $M$.
  We have 
\bea
\sqrt{m_1(t) m_2(t)} &=& e^{-\kappa^\prime t} \sqrt{m_1(0) m_2(0)}  \\
M(t) & = & e^{\xi \sigma_3}  M(0)  e^{\xi \sigma_3} \ . \label{eq:Mt0}
\eea
The overall scale $\sqrt{m_1m_2}$ has a simple dependence on $t$. 
The mixing angle and the (complex) mass ratio evolve
via a transformation that is similar to the effects of the 
familiar seesaw model \cite{seesaw}. 
The difference lies only in that while
$|\xi| \gg 1$ for the seesaw model, for the RGE usually 
$|\xi| \ll 1$ ($\sim 10^{-5}$ for the SM).
Nevertheless, the general analytic solution obtained earlier 
\cite{kuo,kcw} is equally
valid here. We will now discuss this solution in the context of RGE.

  As was shown earlier \cite{kcw}, a symmetric and complex $2\times 2$ 
matrix with ${\rm det}=1$ can be parametrized in a standard way.
We write
\bea
M(t) & = & U(t) e^{-2 \eta \sigma_3} U(t)^T \\
U(t) & = & e^{-i \alpha \sigma_3} e^{-i \theta \sigma_2} e^{-i \phi \sigma_3} 
\ ,
\eea
where $\eta$ is given in terms of the mass eigenvalues by
\be
\eta = {1 \over 4} \ln {m_2 \over m_1} \ .
\ee
At $t=0$, $M(0)$ and $U(0)$ are similarly defined in terms of
$(\eta_0, \theta_0, \alpha_0, \phi_0)$.
Here $m_2$ and $m_1$ are, by definition, positive definite. The relative
phase of the mass eigenvalues is given by $4\phi$ so that $M$ becomes real
for $\phi=0$ and $\phi=\pi/4$, corresponding to the same sign and 
opposite sign mass values, respectively; $\theta$ is the physical mixing 
angle. The phase $\alpha$ can be absorbed into the arbitrary phase of the
charged leptons and is not observable.  
However, as emphasized before (see refs. \cite{Casas} and \cite{kcw}), 
it evolves with t and so influences the evolution of the other parameters.
We can rewrite $M=M(t)$ as \cite{kcw}
\be
M = \left(\ba{cc} 
e^{-2i\alpha}({\rm ch} 2 \bar{\eta} - c_{2\theta}\ {\rm sh} 2\bar{\eta}) &
-s_{2\theta} \ {\rm sh} 2 \bar{\eta} \\
-s_{2\theta} \ {\rm sh} 2 \bar{\eta} & 
e^{2i\alpha}({\rm ch} 2 \bar{\eta} + c_{2\theta} \ {\rm sh} 2\bar{\eta}) \\
\ea \right)
\label{eq:mtk}
\ee
where we use the notation ${\rm ch} 2 \bar{\eta} =\cosh 2 \bar{\eta}$,
$c_{2\theta} = \cos 2\theta$, etc. and $\bar{\eta}=\eta + i \phi$.

 The RGE evolution of the parameters is given by 
$M(0) \rightarrow M(t)$. From Eq.~(\ref{eq:Mt0}), it is obvious that the
off-diagonal elements of $M$ do not evolve.
We have thus the (complex) RGE invariant:
\be \label{eq:inv}
s_{2\theta} \ {\rm sh}2\bar{\eta} = s_{2\theta_0} \ {\rm sh}2\bar{\eta_0} 
\ee
where the subscript 0 denotes the values at the high energy scale $t=0$.
We note that in terms of the physical neutrino masses, 
${\rm sh}2\bar{\eta} 
= (\sqrt{m_2/m_1} e^{i2\phi}-\sqrt{m_1/m_2}e^{-i2\phi})/2$.

 Eq.~(\ref{eq:Mt0}) also implies two complex relations between the diagonal 
elements of $M(t)$ and $M(0)$. They can be used to solve for the four unknowns 
$(\alpha, \theta, \phi, \eta)$ in terms of their initial
values $(\alpha_0, \theta_0, \phi_0, \eta_0)$. After a short 
calculation we find
\bea
\tan 2 \theta & = & {s_{2\theta_0}/(c_{2\Delta\alpha} {\rm ch}2\xi) \over
c_{2\theta_0} -\Sigma_R \tanh 2\xi +\Sigma_I \tan 2\Delta\alpha} \ , 
\label{eq:theta}\\
\tan 2 \Delta\alpha & = & {\Sigma_I \over
       c_{2\theta_0} - \Sigma_R \coth 2\xi } \label{eq:alpha}\\
\coth 2 \bar{\eta_0} & = & \Sigma_R + i \Sigma_I
     ={1 -(m_1/m_2)_0^2 - 2i (m_1/m_2)_0 s_{4\phi_0} \over
      1 +(m_1/m_2)_0^2 - 2 (m_1/m_2)_0 c_{4\phi_0} }   \label{eq:eta}
\eea
where $\Delta \alpha = \alpha - \alpha_0$, the $\Sigma$'s are real,
and the masses in the last equation are at $t=0$. 
These solutions agree with Ref.~\cite{kuo}, where they were obtained 
by a somewhat
different method. We do not write down the solution for $\bar{\eta}$,
since, knowing $\theta$, $\eta$ and $\phi$ can be determined by 
Eq.~(\ref{eq:inv}).

\section{RGE}

The RGEs for the physical parameters can be worked out.
From Eq.~(\ref{eq:Mt0}), the RGE for $M(t)$ is
\be \label{eq:R}
{d \over dt} M =  \chi \{M, \sigma_3\} \ .
\ee
We may decompose $M$ into Pauli matrices,
\be
M=\Sigma^3_{i=0} M_i \sigma_i \ ,
\ee
where the coefficients $M_i$ are complex, and $M_2=0$ since $M$ is symmetric.
Then
\bea
{d \over dt} M_1 & = & 0   \label{eq:M1}\\
{d \over dt} (M_0 \pm M_3) & = & \pm 2 \chi (M_0 \pm M_3)  \ .
\eea
Eq.~(\ref{eq:M1}), which says that the off-diagonal elements of $M$
do not run, is just another form of Eq.~(\ref{eq:inv}),
\be  \label{eq:RG'}
{d \over dt} (s_{2\theta} \ {\rm sh} 2\bar{\eta} ) = 0 \ .
\ee
The diagonal elements of $M$ run according to
\bea
{d \over dt} [ e^{-2i\alpha} ( {\rm ch} 2\bar{\eta} - c_{2\theta} \
    {\rm sh} 2\bar{\eta} )] & = &
 + 2 \chi e^{-2i\alpha} ( {\rm ch} 2\bar{\eta} - c_{2\theta} \
    {\rm sh} 2\bar{\eta} ) \label{eq:diag1}\\
{d \over dt} [ e^{2i\alpha} ( {\rm ch} 2\bar{\eta} + c_{2\theta} \
    {\rm sh} 2\bar{\eta} )] & = &
  - 2 \chi e^{2i\alpha} ( {\rm ch} 2\bar{\eta} + c_{2\theta} \
    {\rm sh} 2\bar{\eta} ) \label{eq:diag2} \ .
\eea
Eqs.~(\ref{eq:diag1}) and (\ref{eq:diag2}) can be readily rewritten in
terms of RGE for the various parameters. It is found that, for the
unobservable phase $\alpha$,
\be   \label{eq:dalpha}
{d \over dt}  \alpha = \chi \ {s_{4\phi} \over {\rm sh} 4\eta}
\ee
The RGEs for the physical parameters are
\bea
{d\eta \over dt} & = & - \chi \ c_{2\theta} \label{eq:deta} \\
{d\phi \over dt} & = & - \chi \ c_{2\theta} \
         {s_{4\phi} \over {\rm sh} 4 \eta}  \label{eq:dphi} \\
{d\theta \over dt} & = & \chi \  s_{2\theta} \ [
        c^2_{2\phi} \coth 2 \eta + s^2_{2\phi} \tanh 2\eta]  \ .
      \label{eq:dtheta}
\eea
It is easy to verify that Eq.~(\ref{eq:RG'}) is satisfied.
Also, the last equation
agrees with the known result for the case of real mass matrix 
\cite{babu,Casas,other}
\be
{d\theta \over dt} = \chi s_{2\theta} { m_2 + m_1 \over m_2 - m_1}
\ee
valid for same sign ($\phi=0$) and opposite sign ($\phi=\pi/4$, or
$m_1 \rightarrow - m_1$) mass values.
Our result, with the convention of positive definite $m_1$ and $m_2$,
interpolates these two equations.
Note also that the equations can be expressed in terms of
$\tanh 2\bar{\eta}$, with
$[c^2_{2\phi} \coth 2\eta + s^2_{2\phi} \tanh 2\eta] =
1/{\rm Re} (\tanh 2\bar{\eta})$
and
$s_{4\phi}/{\rm sh} 4\eta =
{\rm Im}(\tanh 2\bar{\eta})/{\rm Re} (\tanh 2\bar{\eta})$.
Finally, the right-hand sides of Eqs.~(\ref{eq:dalpha}-\ref{eq:dtheta})
are independent of the unobservable phase $\alpha$, as it should be.
Explicitly, from Eq.~(\ref{eq:Mt0}), $e^{\xi \sigma_3}$ commutes with
$e^{i\alpha \sigma_3}$, so that any RGE evolution does not depend on
$\alpha$.

  It is interesting to note that, for $\phi=0$, there is a simple
geometrical interpretation for Eqs.~(\ref{eq:deta}) and (\ref{eq:dtheta}).
As was shown earlier, by writing 
$M(t+dt)=e^{\xi \sigma_3}e^{-2\eta \vec{\sigma} \cdot \hat{n}_{2\theta} }
e^{\xi \sigma_3}$ with $\xi=\chi dt$, Eq.~(\ref{eq:Mt0}) can be regarded
as a velocity addition problem ($\delta \vec{v} + \vec{v} + \delta \vec{v}$)
with $\vec{v}=(\tanh 2\eta) \hat{n}_{2\theta}$ and 
$\delta \vec{v}= - \chi dt \hat{n}_3$.
Here $\hat{n}_{2\theta}=c_{2\theta} \hat{n}_3 + s_{2\theta} \hat{n}_1$.
The resultant Lorentz boost has the rapidity $2(\eta + d\eta)$, with
$d\eta =\delta v_{||}=-\chi dt c_{2\theta}$,
while its angle with respect to the third axis is increased by
$2d\theta=2\delta v_{\perp}/|\vec{v}| = 
2 \chi dt s_{2\theta}/\tanh 2 \eta$.

\section{Phase Portraits}

To better understand the qualitative effects of the RGE,
we rewrite Eqs.~(\ref{eq:deta}-\ref{eq:dtheta}) in terms of a slightly 
different combination of the physical mass eigenvalues \cite{Casas}
\be
z = \coth 2 {\bar \eta} 
  = { m_2 e^{2i\phi} + m_1 e^{-2i\phi} \over m_2 e^{2i\phi} - m_1 e^{-2i\phi} }
\ee
In terms of this variable, the RGEs are
\bea
{d z \over dt} & = & 4 \chi \ c_{2\theta} { (-1 + z^2) z^* \over z + z^*} 
\label{eq:rgezz} \\
{d\theta \over dt} & = & 2 \chi \  s_{2\theta} { | z |^2 \over z + z^*}
\label{eq:rgez}
\eea
where $z^*$ is the complex conjugate of $z$.

The neutrino mass matrix, Eq.~(\ref{eq:mtk}),  
possesses various symmetries.  For example, under the transformation
\be
\theta \rightarrow {\pi \over 2} - \theta \\
\label{eq:sym1}
\ee
accompanied by an interchange of the mass eigenvalues
\bea
\eta & \rightarrow & - \eta \ \  ( m_2 \leftrightarrow m_1 ) \\
\phi & \rightarrow & - \phi
\label{eq:sym2}
\eea
or, equivalently,
\be
z \rightarrow -z 
\label{eq:sym3}
\ee
the diagonal elements are invariant while the off-diagonal elements 
change sign.
However the sign of the off-diagonal mass-matrix element,
which is the RGE invariant (Eq.~(\ref{eq:inv})),
is not a physical observable since it may be absorbed
in the unphysical phases with a redefinition of the neutrino wave function
$(\nu_\mu , \nu_\tau) \rightarrow (\nu_\mu ,- \nu_\tau)$.
Thus the mass matrix and the RGE evolution equations respect this symmetry.

The evolution equations in terms of complex $z$ and $\theta$, 
Eqs.~(\ref{eq:rgezz}) and (\ref{eq:rgez}), 
have five fixed points.  
The evolution is always towards or away from one of these fixed points,
i.e.~the parameters do not evolve towards infinity.
Of the five fixed points, only three are physically 
distinct in that 
the above mass matrix symmetry maps two fixed points into two others.
We have obtained the stability of the fixed points for t {\it decreasing }
(i.e.~running towards the infrared) in the SM by finding the eigenvalues
of the Jacobian (see e.g.~\cite{strogatz}).
The attractive fixed points are $\theta = 0 , z = +1$ 
and $\theta = \pi/2 , z = -1$ which corresponds to no mixing and a 
massless muon-neutrino.
The repulsive fixed points are $\theta = 0 , z = -1$ and 
$\theta = \pi/2 , z = +1$ which corresponds
to no mixing and a massless tau-neutrino.
The final fixed point is a saddle point,
attractive in some directions and repulsive
in others, and it is at $\theta = \pi/4 , z = 0$.
This point corresponds to maximal mixing with equal magnitude 
but opposite sign mass eigenvalues.
These stabilities are for the SM, for the MSSM all evolution directions
are reversed, so the attractors and repulsors are reversed also.

\begin{figure}[t]
\centerline{\epsfysize=20.cm\epsfbox{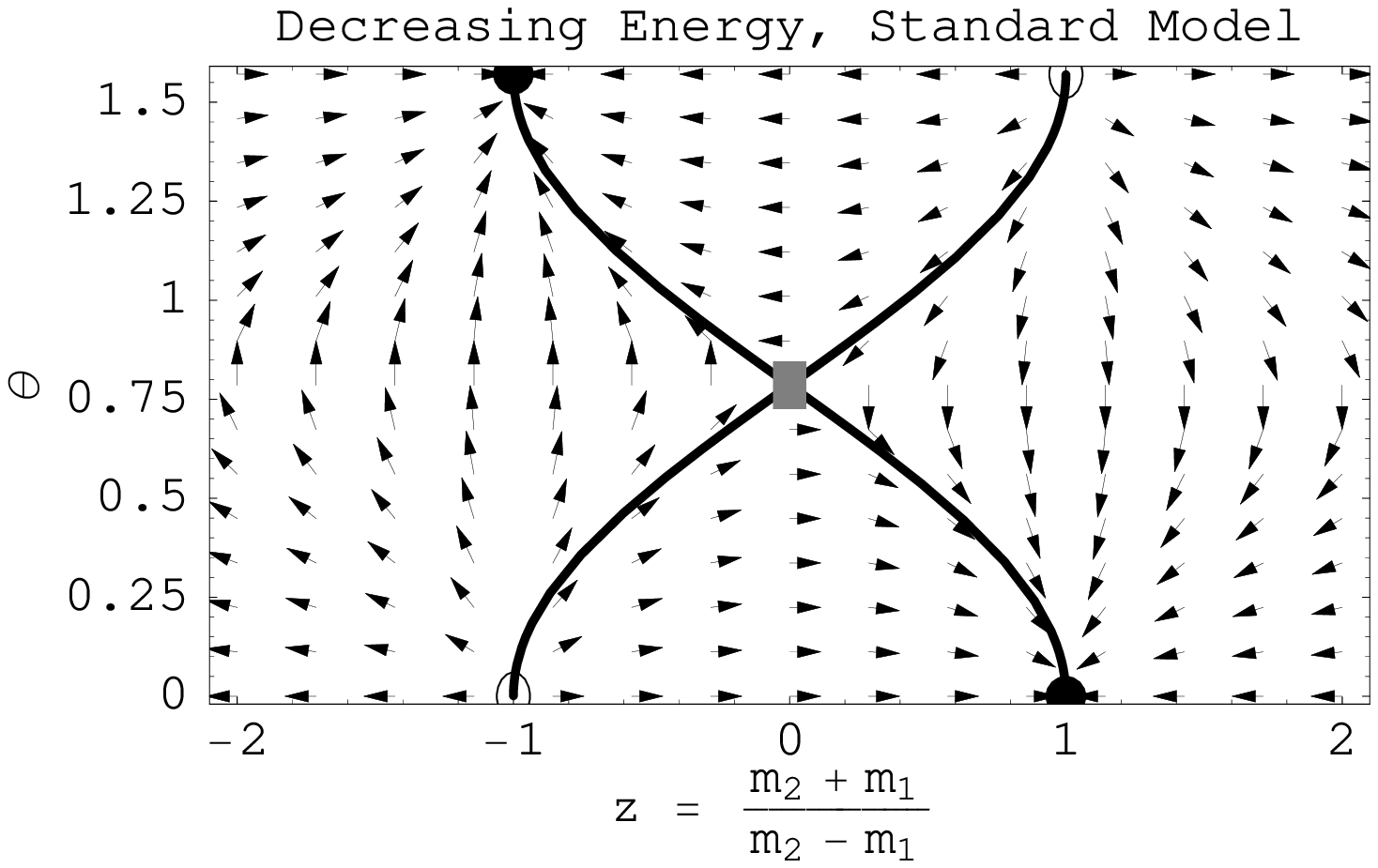}}
\caption[] {
Phase portrait for the Standard Model RG equations when Im($z$) = 0. 
The different fixed points are shown with solid circles, 
open circles and grey square denoting attractors, 
repellors and saddle point. 
The solid curve shows the trajectories that connect the fixed points.
The arrows are reversed for the MSSM.
\label{fig3} }
\end{figure}

\begin{figure}[t]
\centerline{\epsfysize=20.cm\epsfbox{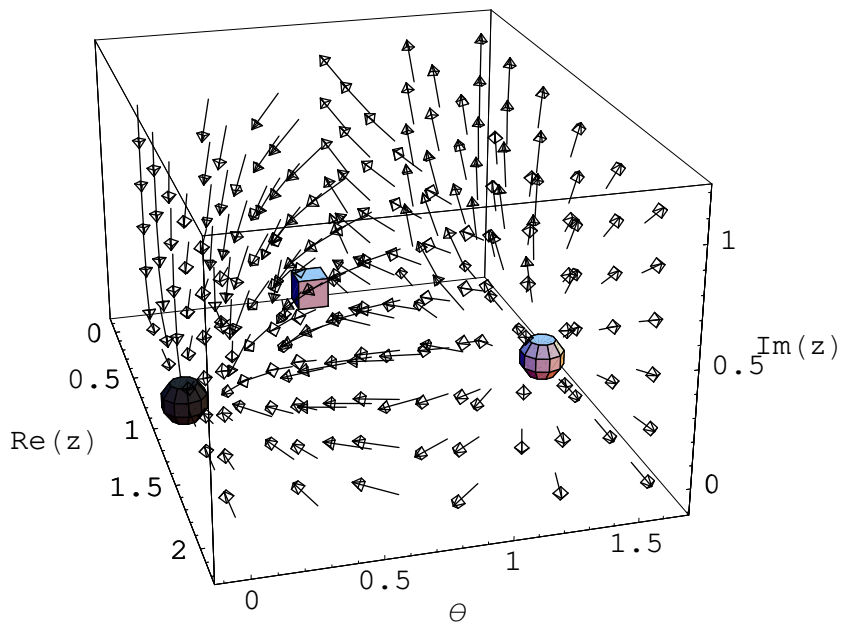}}
\caption[] {
The SM RGE evolution in Re($z$), $\theta$, Im($z$) space, 
for the range $0 < $ Re($z$), $0 < $ Im($z$), $0 < \theta < \pi/2$. 
The different fixed points are all confined to the 
${\rm Im}(z)=0$ plane.  They 
are shown using a dark sphere, light sphere and a cuboid
which denote the attractor, repellor and saddle point.
\label{fig4} }
\end{figure}

\begin{figure}[t]
\centerline{\epsfysize=20.cm\epsfbox{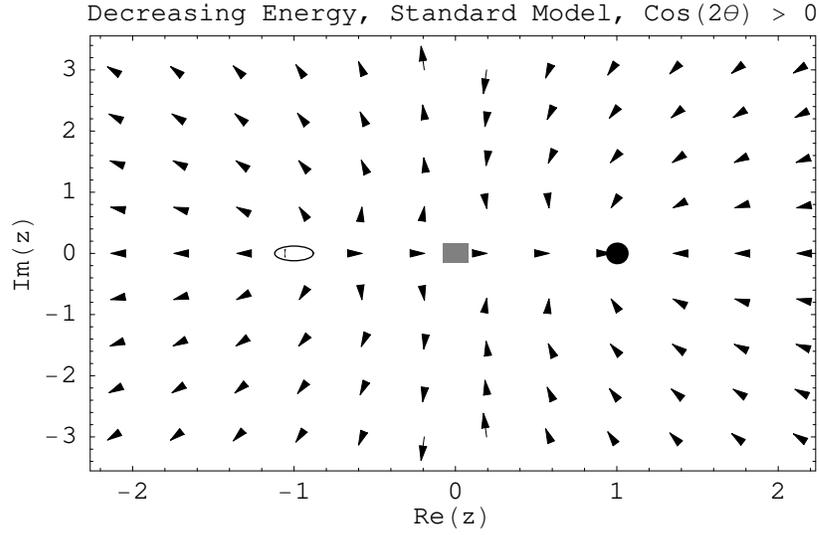}}
\caption[] {
The SM RGE evolution in the Re($z$), Im($z$) plane for $\cos 2\theta > 0$. 
The $\theta=0$ fixed points are shown with solid circle, 
open circle and grey square denoting attractor, repellor and saddle point. 
For $\cos 2\theta < 0$, the figure is similar except all directions 
(and stabilities) are reversed. Note that $\theta$ also evolves.
\label{fig5} }
\end{figure}

These RGEs and the fixed points are graphically displayed in 
Figs.~\ref{fig3}-\ref{fig5} 
where the direction fields are plotted for t decreasing. 
Starting from some initial point specified by the high energy theory,
the evolution of $z$ and $\theta$ follow a trajectory in these spaces
and the plotted unit arrows show the directions tangent to this trajectory.
The direction field is independent of t and $\chi$. 
The trajectories can be found by taking the RGE equations,
Eqs.~(\ref{eq:rgezz}) and (\ref{eq:rgez}), dividing one by the other 
and integrating to get
\be
{\sin^2 2 \theta  \over z^2 - 1} =  \zeta 
\label{eq:zeta}
\ee
where $\zeta$ is a constant.  
This equation is just the square of the RGE invariant, Eq.~(\ref{eq:inv}).

If the mass matrix is initially real, then RGE evolution preserves this.
The evolution of real parameters, Im($z$) = 0, is shown in Fig.~\ref{fig3}.
The symmetries of the neutrino mass matrix, 
Eqs.~(\ref{eq:sym1}) and(\ref{eq:sym3}),
are apparent.
The dark curves in Fig.~\ref{fig3} correspond to  
$\zeta = - 1$ in Eq. (\ref{eq:zeta}).
These curves connect all of the fixed points.
Note that half of these dark curves flow towards the maximal mixing fixed 
point and end there, the only trajectory that does so. 
Trajectories with $\zeta < -1$ or $\zeta \ge 0$ 
are those on the left and right of the plot
and they can pass through maximal mixing.
Trajectories with $-1 < \zeta < 0$ connect positive and negative $z$ values 
and never attain maximal mixing.

The general behavior for Im($z$)$\neq 0$ is shown in Fig.~\ref{fig4}, 
where the vector field is plotted in the full, 3-D parameter space.
This figure is for the range $0 < {\rm Re}(z)$, $0 < {\rm Im}(z)$, 
$0 < \theta < \pi/2$,
so the `floor' of this figure corresponds to the right half of 
Fig.~\ref{fig3}.  The behavior in other regions of the parameter space 
can be obtained from this by symmetry operations.
The three fixed points are shown using a dark sphere, 
light sphere and a cuboid to denote the attractor, repellor and saddle point. 
Note that the figure resembles the vector field of a dipole,
with evolution away from the repellor and towards the attractor.

The evolution equations are singular at Re($z$) = 0 if Im($z$)$\neq 0$.
Thus large changes in the mixing angle are possible then, despite
the small size of $\chi$.
The behavior in this region is shown in Fig.~\ref{fig5}. Here 
the direction field is plotted for the Im($z$), Re($z$) plane.
Note that this direction field is independent of the magnitude of $\theta$,
except that the direction of flow does depend on the sign of $\cos 2\theta$.
The fixed points for $\theta = 0$ and $\theta= \pi/4$ are
included in the plot. 
As Fig.~\ref{fig5} shows, the flow across the Re($z$) = 0 value does not occur
if ${\rm Im}(z) \neq 0$.  The flow on the right-hand side is towards 
the stable
fixed point there.  On the left-hand side, the flow is away from the repulsor,
but at the same time the mixing angle is increasing and eventually 
exceeds $\theta=\pi/4$
at which point the flow direction reverses, the repulsor becomes an attractor
and the evolution approaches the now stable fixed point 
by retracing its path in the Im($z$), Re($z$) plane.

\section{Analysis}

We turn now to a quantitative analysis of our results.
First, we consider the case of real mass matrices, 
$\phi_0 = 0$ or $\pi/4$, so that $\Delta\alpha=0$ from Eq.~(\ref{eq:alpha})
and 
\be  \label{eq:tan2theta}
\tan 2\theta = { s_{2\theta_0}/{\rm ch}2\xi \over 
       c_{2\theta_0} - \Sigma_R \tanh 2\xi}
\ee
where, from Eq.~(\ref{eq:eta})
\be   \label{eq:sigR}
\Sigma_R = \left\{ \ba{cl}
 \left({m_2 + m_1 \over m_2 - m_1}\right)_0 = \coth 2\eta_0 \ , & \phi_0=0 \\
 \left({m_2 - m_1 \over m_2 + m_1}\right)_0 = 
\tanh 2\eta_0 \ , & \phi_0=\pi/4 \ .
        \ea \right .
\ee
Since $|\xi| \ll 1$, ($|\xi| \sim 10^{-5}$ for the SM and 
$|\xi| \le {\cal O}(10^{-2})$ for the MSSM), we see immediately that
for $\phi_0=\pi/4$ (opposite sign mass eigenvalues), $\theta \simeq \theta_0$.
For $\phi_0=0$ (same sign masses), $\tan 2\theta$ exhibits the resonance
behavior, and maximal mixing is obtained if
\be
c_{2\theta_0} = \tanh 2\xi/\tanh 2\eta_0 \ , \;\;\;\;  (\phi_0=0) \ .
\ee
For a generic value of $\theta_0 < \pi/4$, this condition can be
fulfilled provided that $\xi$ and $\eta_0$ are of the same sign, and that
$\eta_0$ is very small (nearly degenerate masses).
In terms of the masses, if we define
\be
\delta_0 =1 - (m_1/m_2)_0
\ee
then the resonance condition becomes
\be \label{eq:res}
c_{2\theta_0} \simeq 4 \xi/\delta_0 
\hskip 0.5in (\phi_0=0) \ .
\ee
 For SM, $\xi<0$, it is seen that $\theta < \theta_0$, if $m_2 > m_1$.
That is, under "normal" situations, RGE drives the mixing angle toward
zero as the energy decreases.
For MSSM, $\xi >0$, we see that $\theta$ can become large if there is
near degeneracy, $m_2 \simeq m_1$ ($m_2 > m_1$).
Of course, if $\theta_0 > \pi/4$ (which amounts to the interchange
$m_1 \leftrightarrow m_2$), or if $m_1 > m_2$, 
the direction of the running of the mixing angle is reversed.
These conclusions agree with earlier results obtained by numerical 
integrations of the RGE\footnote{Note also that, for general values of $\phi$,
similar behaviors can be deduced from Eq.~(\ref{eq:dtheta})}.

 At the same time, the RGE invariant (Eq.~(\ref{eq:inv})) gives the running 
of the mass ratio
\bea
s_{2\theta} \ {\rm sh} 2\eta & = & s_{2\theta_0} \ {\rm sh} 2 \eta_0 \ , 
   \;\;\;\; (\phi_0=0)  \\
s_{2\theta} \ {\rm ch} 2\eta & = & s_{2\theta_0} \ {\rm ch} 2 \eta_0 \ , 
   \;\;\;\; (\phi_0=\pi/4)  \ .
\eea
While $\eta \sim \eta_0$ for $\phi_0 = \pi/4$, for $\phi_0=0$ and at the
resonance, $s_{2\theta} \gg s_{2\theta_0}$, it is seen that
${\rm sh} 2 \eta \ll {\rm sh} 2 \eta_0$,  {\it i.e.}, the masses are driven
toward even more degeneracy as $\theta$ becomes maximal from 
a small $\theta_0$.

\begin{figure}[t]
\centerline{\epsfysize=20.cm\epsfbox{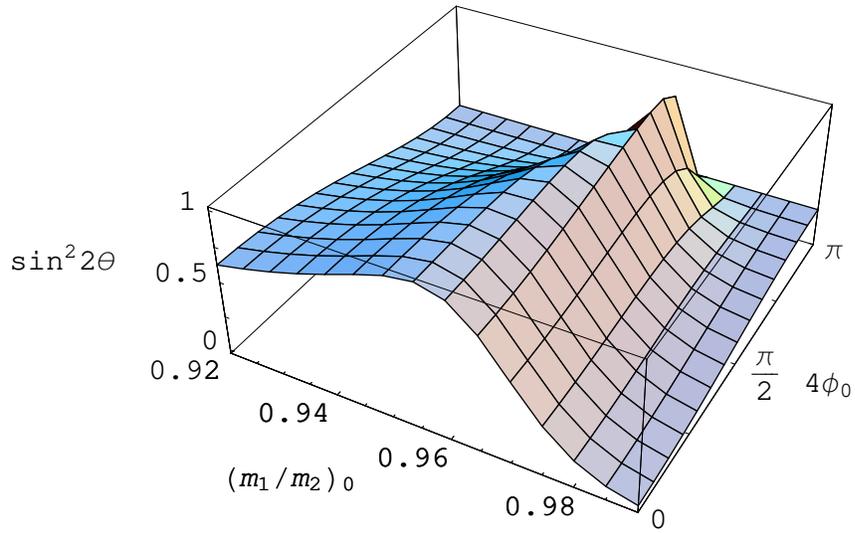}} 
\caption[] { 3D plot for $\sin^2 2\theta$ vs. initial values
$(m_1/m_2)_0$ and $4\phi_0$. The inputs are fixed at 
$\theta_0=\pi/12$ and the RGE factor $\xi=0.01$ (MSSM). 
\label{fig1} } 
\end{figure}

\begin{figure}[t]
\centerline{\epsfysize=20.cm\epsfbox{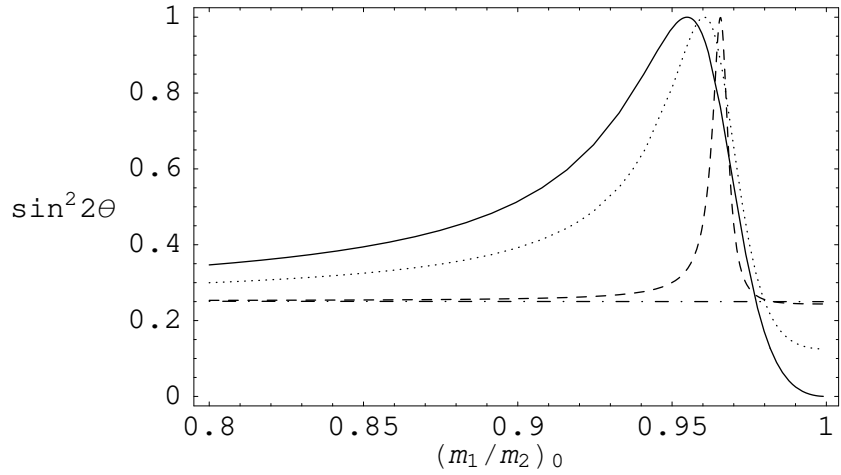}} 
\caption[] {$\sin^2 2\theta$ vs. $(m_1/m_2)_0$ with inputs
$\theta_0=\pi/12$ and $\xi=0.01$ (MSSM). The solid, dotted, dashed, and 
dot-dashed curves 
correspond to $4\phi_0=0, \pi/2, 0.9 \pi$ and $\pi$ respectively.  
\label{fig2} }
\end{figure}

  For complex mass matrices ($\phi_0 \neq 0$ or $\pi/4$), we need to use
Eq.~(\ref{eq:theta}). 
From Eq.~(\ref{eq:dtheta}), it is seen that for $\chi \ll 1$, 
$d\theta/dt$ is appreciable only if $\eta \ll 1$. In other words,
RGE effect can be large only if $\delta= 1 -(m_1/m_2) \ll 1$, as for real 
matrices.  We will confine our discussion in this parameter region
$\delta \ll 1$. Further, for definiteness, we assume that
$\xi >0$ and $\eta_0 > 0$. (The other possibilities can be easily analyzed.)
For finite values of $\phi_0$, it is easy to see that
\bea
\Sigma_R & \simeq & \delta_0/(2 s^2_{2\phi_0} ) \\
\Sigma_I & \simeq & - \cot 2\phi_0
\eea
where higher order terms in $\delta_0$ are neglected. 
The condition for maximal mixing is then
\be \label{eq:R2}
{\delta_0 \over 4 \xi} \simeq (c_{2\theta_0} \ s^2_{2\phi_0}
                           + c^2_{2\phi_0}/c_{2\theta_0} ) \ ,
       \;\;\;\;   (\phi_0 \neq 0, \pi/4)
\ee
This reduces to Eq.~(\ref{eq:res}) as $\phi_0 \rightarrow 0$. However,
when $\phi_0 = \pi/4$, we had found earlier that there is no resonance.
This means that the limit $\phi_0 \rightarrow \pi/4$ is actually 
very delicate. As we will see, 
what happens is that the width of the resonance also shrinks to zero
as $\phi_0 \rightarrow \pi/4$. 
This behavior is borne
out in the numerical calculation shown in Figs.~\ref{fig1} and \ref{fig2}.

   Although the analytic formulae for $\theta$, $\eta$ and $\phi$
have been explicitly given, as we have seen in the previous analyses,
their detailed behaviors are quite intriguing. It is useful to have an 
overview of these functions. For this purpose we present, in Figs.~\ref{fig1} 
and \ref{fig2}, numerical calculations of the function $\theta(t)$.
Traditionally the plot of $\theta(t)$ is made with $t$ as the variable.
However, as shown in Eqs.~(\ref{eq:res}) and (\ref{eq:R2}), the 
interesting region is determined by $\delta_0/\xi \sim {\cal O}(1)$.
So an equivalent way is to plot $\theta$ vs. $\delta_0$ 
(or $(m_1/m_2)_0$), while
keeping $t$ as a constant parameter.
In Fig.~\ref{fig1}, we present a 3D plot of $\sin^2 2\theta$ vs. 
$(m_1/m_2)_0$ and 
$4\phi_0$. We take a positive $\xi=\chi t=10^{-2}$ (corresponding to MSSM
with $\tan \beta = 40$ and $t={1\over 16 \pi^2} \ln
(m_Z/M_X)$, $M_X\simeq 3\times 10^{10}$ GeV).
This sets the scale in $(m_1/m_2)_0$, with 
$(1-m_1/m_2)_0 =\delta_0 \sim 10^{-2}$.  
Thus for a given initial value $\theta_0=\pi/12$, the plot gives
$\sin^2 2\theta$ as a function of the initial parameters 
$(m_1/m_2)_0$ and its phase
$4\phi_0$. The resonant behavior of $\sin^2 2\theta$ is obvious.
The position of the maximum is, for $\phi_0=0$, 
at $\delta_0 \simeq 8 \xi/\sqrt{3}$.
The position of the maximum shifts with changing $\phi_0$, and is
well described by Eq.~(\ref{eq:R2}).

  Fig.~\ref{fig2} shows 2D slices of the 3D plot in Fig.~\ref{fig1}.
We plot, for various values of $\phi_0$, $\sin^2 2 \theta$ vs. $(m_1/m_2)_0$. 
It shows clearly the shifting positions of the maxima as $\phi_0$ changes.
Also notice the change of width as a function of $\phi_0$. 
It can be shown that, provided $4 \xi^2 \ll c^2_{2\theta_0} 
\tan^2 2\phi_0$, the width is given by 
\be
\Delta (m_1/m_2)_0 \simeq 8 \xi  c_{2\phi_0}  (s_{2\theta_0}/c^2_{2\theta_0}) 
 \sqrt{c^2_{2\phi_0}  + c^2_{2\theta_0} s^2_{2\phi_0} } 
\ ,
\ee
so that the resonance disappears as $\phi_0 \rightarrow \pi/4$,
in agreement with Eqs.~(\ref{eq:tan2theta}) and (\ref{eq:sigR}). 
Finally, the 
singular nature of the limit $\phi_0 \rightarrow \pi/4$ is confirmed when
we compare the curves with $\phi_0=0.9\pi/4$ and $\phi_0=\pi/4$.

  We do not show 3D plots for $\eta$ and $\phi$. Since from the RGE
invariant,
\bea
s_{2\theta} \ {\rm sh} 2\eta \ c_{2\phi} & = & 
s_{2\theta_0}  \ {\rm sh}2\eta_0 \ c_{2\phi_0} \ , \label{eq:inv1} \\
s_{2\theta} \ {\rm ch} 2\eta \ s_{2\phi} & = & 
s_{2\theta_0}  \ {\rm ch}2\eta_0 \ s_{2\phi_0} \ , \label{eq:inv2}
\eea
once we know $\theta$, they can be easily determined.
Note that for nearly degenerate masses ($\eta \ll 1$), it can be 
seen from Eq.~(\ref{eq:inv2}) that there is an
anti-correlation between $\theta$ and $\phi$ with the phase $\phi$ being
minimal at maximal mixing.
Also, at maximal mixing and with nearly degenerate masses, 
simple expressions for the neutrino mass ratio and
the relative phase can be deduced from Eqs.~(\ref{eq:inv1}) and 
(\ref{eq:inv2}),
\bea
s_{2\phi} & \simeq &  s_{2\theta_0} s_{2\phi_0} \\
\delta & \simeq & \delta_0
    s_{2\theta_0} c_{2\phi_0} / 
     \sqrt{1 - s^2_{2\theta_0} s^2_{2\phi_0} } \ , \label{eq:mratio}
\eea
where $\delta = 1 -m_1/m_2$.
As can be seen from Eq.~(\ref{eq:mratio}), for small $\theta_0$,
the neutrino masses become more degenerate at maximal mixing.

\section{Conclusions}

We have examined the RGEs for the physical, two-flavor, neutrino 
parameters: mixing angle, 
mass ratio and their relative phase in the SM and MSSM.
These equations turn out to have relatively simple forms which we 
analyze in detail.
The qualitative nature of the evolution is clearly illustrated in 
our vector field plots.
A more quantitative description is also given using our exact, 
analytical solution for the evolution. 
In addition, we have also found a complex RGE invariant which 
correlates the running of the three physical parameters.

The phase portraits show the direction of the evolution throughout 
the parameter space.
The fixed points and their stability are included on the plots. 
It should be noted that maximal mixing ($\theta=\pi/4$) is a saddle point
in both the SM and MSSM,
and the infrared stable point (attractor) in the SM (MSSM)
corresponds to no mixing and a massless muon-neutrino (tau-neutrino).
Thus to get maximal mixing at the experimental scale requires
a particular choice of mass ratio and phase at the high-energy scale.

The parameter choices that give maximal mixing have been calculated with
our exact solution.  As is well-known for real parameters,
large evolution requires nearly degenerate masses.
With the addition of a complex phase, the peak position
shifts and the resonance region becomes
narrower, so to achieve maximal mixing at the experimental scale 
requires a finer tuning of parameters at the high-energy scale.

  In the literature, considerable interest has been focused on the possibility
of generating large mixing through RGE running.  The detailed solution for
RGE suggests both opportunities and limitations for this scenario.
Since the end result depends very sensitively on the initial conditions, 
any such model must be treated carefully.  In particular, one needs to study
the three flavor problem in detail.
Although it is very difficult to obtain exact solutions in this case, 
approximate solutions seem well within reach. We hope to return to this
problem in the future.

\bigskip

 {\bf Acknowledgements:} \ 
T. K. K. and G.-H. W. are supported in part by DOE grant No.
DE-FG02-91ER40681 and No. DE-FG03-96ER40969, respectively.
J.P. is supported by NSF grant PHY-0070527.



\end{document}